\documentclass[12pt]{article}
\usepackage{graphicx}

\newcommand{\beq}{\begin{equation}}
\newcommand{\eeq}{\end{equation}}
\newcommand{\beqn}{\begin{eqnarray}}
\newcommand{\eeqn}{\end{eqnarray}}
\newcommand{\beqns}{\begin{eqnarray*}}
\newcommand{\eeqns}{\end{eqnarray*}}

\begin{document}

\begin{titlepage}
\begin{center}

\hfill USTC-ICTS/PCFT-22-25\\

\vspace{2.5cm}

{\large {\bf Note on rare $Z$-boson decays to double heavy quarkonia}}\\
\vspace*{1.0cm}
 {Dao-Neng Gao$^\dagger$ and Xi Gong$^\ddagger$\vspace*{0.3cm} \\
{\it\small Interdisciplinary Center for Theoretical Study,
University of Science and Technology of China, Hefei, Anhui 230026
China}\\
{\it\small Peng Huanwu Center for Fundamental Theory, Hefei, Anhui 230026 China}}

\vspace*{1cm}
\end{center}
\begin{abstract}
\noindent
Within the standard model, we have investigated rare $Z$-boson decays into double heavy quarkonia, $Z\to VV$ and $Z\to VP$ with $V$ denoting vector and $P$ denoting pseudoscalar quarkonia, respectively. It is assumed that the leading-order QCD diagrams would give the dominant contributions to these processes, and the corresponding branching fractions, for instance, ${\cal B}(Z\to J/\Psi J/\Psi)$ has been estimated to be around $10^{-13}$ in the literature. However, these decays could also happen through the electromagnetic transition $Z\to V\gamma^*$ and $Z\to P\gamma^*$, with the virtual photon transforming into $V$. Interestingly, the smallness of the vector quarkonium mass can give rise to a large factor $m_Z^2/m_V^2$, relative to the QCD contributions, which thus counteracts the suppression from the electromagnetic coupling. We systematically include these two types of contributions in our calculation to predict branching fractions for these decays. Particularly, due to the virtual photon effects, it is found that ${\cal B}(Z\to J/\Psi J/\Psi)$ will be significantly enhanced, which could be up to $10^{-10}$.
\end{abstract}

\vfill
\noindent
$^{\dagger}$ E-mail address:~gaodn@ustc.edu.cn\\\noindent
$^{\ddagger}$ E-mail address:~gonff@mail.ustc.edu.cn
\end{titlepage}

\section{Introduction}

The large rate of $Z$ boson production at the LHC may facilitate the experimental analysis of rare $Z$-boson decay channels. In 2019, a search for rare $Z$-boson decays into a pair of heavy vector quarkonia, $Z \to V V $ ($V=J/\Psi, \Upsilon$), has been firstly performed by the CMS Collaboration \cite{CMS19}, and upper limits on the branching fractions have been obtained.
Very recently, these upper limits have been updated in Ref. \cite{CMS22} as
\beq\label{brJPsi}
{\cal B}(Z\to J/\Psi J/\Psi) < 11\times 10^{-7}
\eeq
and
\beq\label{brUpsilon}
{\cal B}(Z\to \Upsilon(1S) \Upsilon(1S)) < 1.8\times 10^{-6}
\eeq
at the 95\% confidence level, respectively.

In the standard model (SM), rare decays of $Z\to VV$ have already been calculated in Refs. \cite{BR90, LL18}. It is generally believed that the lowest QCD diagrams, as displayed in Figure 1, would give rise to the dominant contributions to these transitions in the SM. In this paper, we will reexamine the analysis of these decays in the SM. It will be shown below that, besides the diagrams in Figure 1, some other diagrams, as displayed in Figure 2, may also bring about important contributions due to the virtual photon exchange. Therefore it is necessary to carry out a systematical calculation of the branching ratios of $Z\to VV$ decays by including all of the relevant diagrams, which will be helpful for the future study to compare the SM predictions with experimental measurements.

The paper is organized as follows. In Section 2, we update the leading-order QCD analysis of $Z\to VV$. The virtual photon contributions to $Z$-boson decays into the same fianl states will be studied in Section 3. In Section 4, rare $Z\to V P$ modes with $P$ denoting pseudoscalar heavy quarkonia is similarly analyzed. Finally, we summarize our results in Section 5.

\section{Leading-order QCD contributions to $Z\to VV$}

Let us first deal with Figure 1, which gives the leading-order QCD contributions to $Z\to VV$ transitions. To explicitly evaluate the decay amplitudes, one should project $Q\bar{Q}$ into the corresponding hadron states. As a reasonable approximation for the leading order calculation, in the present work we adopt the nonrelativistic color-singlet model \cite{colorsinglet}, in which the quark momentum and mass are taken to be one half of the corresponding quarkonium momentum $p$ and mass $m_V$, i.e. $p_Q=p_{\bar{Q}}=p/2$ and $m_{V}=2m_Q$.  Thus for the $Q\bar{Q}$ pair to form the heavy quarkonium $V$, one can replace the combination of the Dirac spinors for $Q$ and $\bar{Q}$ by the following projection operator \cite{barger, Jia2007}
\beq\label{projector}
{v(p_{\bar{Q}})} {\bar{u}(p_Q)} \longrightarrow \frac{\psi_{V}(0)I_c}{2\sqrt{3m_V}} {\epsilon\!\!/}^*(p\!\!\!/+m_V),
\eeq
where $I_c$ is the $3\times 3$ unit matrix in color space and $\epsilon^{*\mu}$ is the polarization vector of the heavy quarkonium $V$. $\psi_{V}(0)$ is the wave function at the origin for $V$, which is a nonperturbative parameter.

Using the standard $Z$-boson and gluon couplings to quark pair, one can perform the direct calculation from Figure 1, which gives
\beq\label{amp1}
{\cal M}_1=\frac{256\pi g^Q_a g \alpha_s m_V^2}{3\cos\theta_W m_Z^4}\left(\frac{\psi_V(0)}{\sqrt{m_V}}\right)^2\varepsilon^{\alpha\beta\mu\nu}\epsilon^*_\alpha(q) \epsilon^*_\beta(p)\epsilon^Z_\mu(k) (p-q)_\nu.
\eeq
Here we take the two quarkonia in the final state, and $Z$-boson momenta to be $p$, $q$, and $k=p+q$, respectively.  $g$ is the weak SU(2)$_L$ coupling constant,$\theta_W$ is the Weinberg angle, $g_a^Q$ is the axial-vector coupling of the $Z$ to the quark $Q$, and $g_a^Q=T_3^Q$ with $T_3^Q$ denoting the third component of the weak isospin of the heavy quark. $\alpha_s=g_s^2/4\pi$ and $g_s$ is the strong coupling constant.

\begin{figure}[t]
\begin{center}
\includegraphics[width=8cm,height=3.0cm]{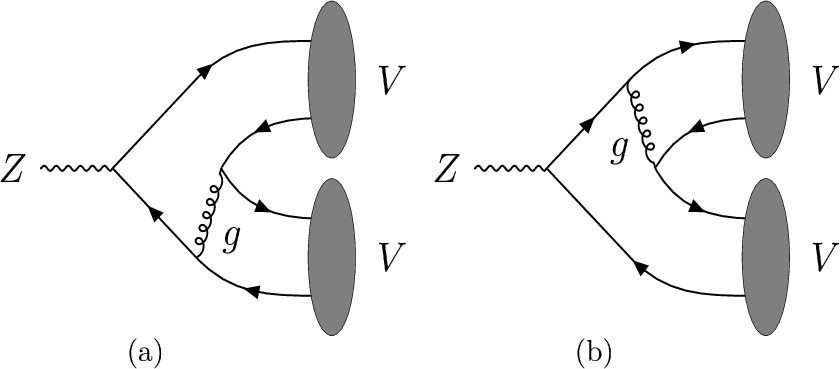}
\end{center}
\caption{Lowest-order QCD diagrams contributing to $Z\to VV$ decays. The solid line with arrow denotes the heavy quarks $Q$ or $\bar{Q}$. Due to the exchange of final identical particles, there are totally four diagrams. }\label{figure1}
\end{figure}

One can find that only $g_a^Q$ appears in the amplitude ${\cal M}_1$, and the vector-component of the $ZQ\bar{Q}$ coupling cannot contribute to $Z\to VV$ decays due to the charge conjugate invariance\footnote{The charge conjugate invariance is not respected by the weak interaction since $Z$-boson couplings to quarks involve both vector current and axial-vector current, which have the different $C$-parity. On the other hand, these currents hadronize into the final heavy quarkonia via the strong or electromagnetic interactions, as shown in Figures 1 and 2, which should obey the charge conjugate symmetry.}. This has also been shown explicitly in Ref. \cite{BR90}. After squaring the amplitude and summing/averaging over the polarizations of the final or initial particles, one can obtain the decay rate as
\beq\label{rateQCD}
\Gamma(Z\to VV)=\frac{512\pi g^2\alpha_s^2}{27 \cos^2\theta_W m_Z^5}|\psi_V(0)|^4 \left(1-\frac{4m_V^2}{m_Z^2}\right)^{5/2}.
\eeq
Note that $\Gamma(Z\to VV)$ will not vanish when we set $m_V=0$. This seems to be in contrast with the Landau-Yang theorem \cite{Landau-Yang}, which states that a massive vector  like $Z$-boson cannot decay into two one-shell photons. Here the difference is that the final vector particle $V$ is also massive, and it has the longitudinal polarization. In the limit of $m_V\to 0$, the longitudinal component will be proportional to $1/m_V$. One can easily find that, from eq. (\ref{amp1}), when one of the final vector boson is longitudinally polarized, the $m_V$ dependence of the amplitude will disappear. Thus, numerically, by taking
\beqn\label{psiJY}
\psi^2_{J/\psi}(0)=0.073^{+0.011}_{-0.009} {\rm GeV}^3,\nonumber\\\\
\psi^2_\Upsilon(0)=0.512^{+0.035}_{-0.032}{\rm GeV}^3\nonumber
\eeqn
from Refs. \cite{BCKLY08, CLY11, BPSV13} and $\alpha_s(m_Z)=0.118$, we have
\beq\label{brJQCD}{\cal B}(Z\to J/\Psi J/\Psi) = 1.5\times 10^{-13}\eeq
and
\beq\label{brYQCD}{\cal B}(Z\to  \Upsilon(1S) \Upsilon(1S)) = 6.8 \times 10^{-12},\eeq
which predict small branching fractions for these processes.

Note that our analytic expression for $\Gamma(Z\to VV)$ in Eq. (\ref{rateQCD}) will be identical to the one shown in eq. (3) of Ref. \cite{BR90} if we take
$$\psi_V(0)\to \sqrt{\frac{1}{4\pi}}R_S(0),\;\;\;\; g^2\to \frac{4\pi \alpha_{\rm em}}{\sin^2\theta_W}$$
with $\alpha_{\rm em}=e^2/4\pi$. However, our numerical results are not in good agreement with predictions obtained in Ref. \cite{BR90}: ${\cal B}(Z\to J/\Psi J/\Psi) = 7.2 \times 10^{-13}$ and ${\cal B}(Z\to  \Upsilon(1S) \Upsilon(1S)) = 6.6 \times 10^{-11}$. Even if the numerical values for $R_S(0)$'s of Ref. \cite{BR90} are used in our calculation, one still cannot reproduce their results\footnote{If we take the standard inputs for $m_Z$,$\Gamma_Z$, $m_{J/\Psi}$, $\sin^2\theta_W$, and $\alpha_{\rm em}$,  together with the values for $R_S(0)$ and the predicted ${\cal B}(Z\to J/\Psi J/\Psi)$ in Ref. \cite{BR90}, we will obtain $\alpha_s\approx 0.33$. One can find that the scale of $\alpha_s\approx 0.33$ is around  $m_{J/\Psi}$ or below. This is obviously not reasonable since, for the leading-order QCD contribution in Figure 1, the virtuality of the gluon is  $O(m_Z^2)$.}.

On the other hand, our prediction for the charmonium mode is larger than the one by the authors of Ref. \cite{LL18}, in which they got ${\cal B}(Z\to J/\Psi J/\Psi)=2.3\times 10^{-14}$ by taking the nonperturbative matrix element $\langle O_1\rangle_{J/\Psi}=0.22$ GeV$^3$ and the strong coupling $\alpha_s(m_Z^2)=0.13$. Further using the relation $\langle O_1\rangle_{J/\Psi}=2 N_C |\psi_{J/\Psi}(0)|^2$ \cite{BCKLY08}, one may find that eq. (\ref{rateQCD}) will give ${\cal B}(Z\to J/\Psi J/\Psi)=4.7\times 10^{-14}$, which is still about factor 2 large.

\section{Contributions to $Z\to VV$ from the virtual photon exchange}

We have updated the analysis of the leading-order QCD contributions to $Z\to VV$ decays. One may argue that the next-to-leading order QCD corrections could be important. This is of course interesting but not the main purpose of the present paper. As pointed out in Introduction, the diagrams in Figure 2 could also lead to significant contributions to these transitions. Thus the $Z$-boson decays into double vector heavy quarkonia could also occur via $Z\to  V \gamma^*$, with the virtual photon transforming into $V$. Similar mechanism has been studied in rare Higgs and other $Z$-boson decays like $h\to \gamma V$ \cite{BPSV13}, $h\to Z V$ \cite{Gao}, $h\to VV$ \cite{GG22}, and $Z\to J/\Psi \ell^+\ell^-$ \cite{Zpsill} processes.

\begin{figure}[t]
\begin{center}
\includegraphics[width=8cm,height=3.0cm]{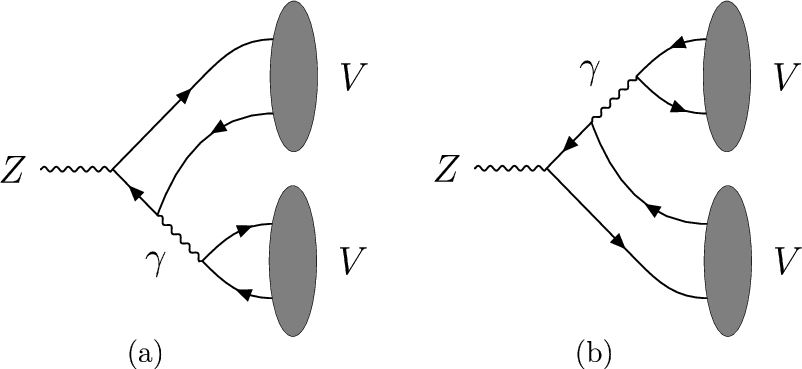}
\end{center}
\caption{Diagrams contributing to $Z\to VV$ decays in the SM through the virtual photon exchange. The solid line with arrow denotes the heavy quarks $Q$ or $\bar{Q}$. Due to the exchange of final identical particles, there are totally four diagrams. }\label{figure2}
\end{figure}

Now from Figure 2, it is straightforward to derive the corresponding decay amplitude for $Z\to VV$, which reads
\beq\label{amp2}
{\cal M}_2=\frac{96\pi g^Q_a g \alpha_{\rm em} e_Q^2}{\cos\theta_W m_Z^2}\left(\frac{\psi_V(0)}{\sqrt{m_V}}\right)^2\varepsilon^{\alpha\beta\mu\nu}\epsilon^*_\alpha(q) \epsilon^*_\beta(p)\epsilon^Z_\mu(k) (q-p)_\nu.
\eeq
 Here $e_Q$ is the electric charge of the heavy quark, $e_c=2/3$ and $e_b=-1/3$. As mentioned above, due to the charge conjugation symmetry, here also only axial-vector component of the $ZQ\bar{Q}$ coupling contributes to the amplitude. One can find that ${\cal M}_1$ and ${\cal M}_2$ have the same structure but different signs. It is generally assumed that ${\cal M}_2$ will be suppressed because of the electromagnetic coupling. However, by computing
\beq\label{ratio}{\cal R}_V=\frac{{\cal M}_2}{{\cal M}_1}=-\frac{9 e_Q^2 \alpha_{\rm em}}{8\alpha_s}\frac{m_Z^2}{m_V^2},\eeq
it is easy to see that the smallness of the vector quarkonia masses will give rise to a large factor $m_Z^2/m_V^2$, which thus counteracts the suppression of $\alpha_{\rm em}$.
Explicitly, we have
\beq\label{r-Jpsi}
{\cal R}_{J/\Psi}=-27.7\eeq
for the charmonium case, and
\beq\label{r-Y}
{\cal R}_{\Upsilon(1S)}=-0.75
\eeq
for the bottomonium case. Here we have used $\alpha_{\rm em}(m_{J/\Psi})=1/132.64$ and $\alpha_{\rm em}(m_{\Upsilon(1S)})=1/131.87$ in the above numerical calculation.
Combining ${\cal M}_1$ and ${\cal M}_2$, we thus obtain the leading order SM predictions for the decay rates of $Z\to VV$ processes as follows
\beq\label{rateJtot}
{\cal B}(Z\to J/\Psi J/\Psi) = (1.1\pm 0.3) \times 10^{-10},
\eeq
and
\beq\label{rateUtot}
{\cal B}(Z\to  \Upsilon(1S) \Upsilon(1S)) = (4.4\pm 0.6) \times 10^{-13}.
\eeq
Here the errors of these results are only due to the uncertainties of $\psi^2_V(0)$'s in eq. (\ref{psiJY}).
Obviously, comparing with the leading-order QCD contributions [eqs. (\ref{brJQCD}) and (\ref{brYQCD})], the large amplitude ${\cal M}_2$ for the charmonioum final states leads to a significant enhancement of ${\cal B}(Z\to J/\Psi J/\Psi)$, which could be up to $10^{-10}$; while the comparable ${\cal M}_1$ and ${\cal M}_2$ with different signs in the bottomonium case results in accidental cancellation in the amplitude, which substantially decreases the decay rate of $Z\to  \Upsilon(1S) \Upsilon(1S)$. In {\it Note added in proof} of their paper, the authors of Ref. \cite{BR90} have pointed out that these decays could be via $Z\to V\gamma^*\to VV$, and the decay rate could be approximated by $\Gamma(Z\to V\gamma){\cal B}(\gamma^*\leftrightarrow V)/2$. Using the results for ${\Gamma}(Z\to V\gamma)$ in Ref. \cite{GKPR80}, they further got ${\cal B}(Z\to J/\Psi J/\Psi)=2.7\times 10^{-11}$ and ${\cal B}(Z\to \Upsilon \Upsilon)=2.1\times 10^{-12}$. In the present paper, we include both types of contributions to predict ${\cal B}(Z\to VV)$.

Only uncertainties of $\psi^2_V(0)$'s are included to evaluate the errors on our results since our present study is performed in the framework of nonrelativistic color-singlet model. On the other hand, one can use the nonrelativistic QCD (NRQCD) factorization method \cite{BBL95} to calculate corrections in powers of $\alpha_s$ and $v$ ($v$ is the heavy-quark velocity in the quarkonium rest frame). Our results are equal to the ones from the NRQCD approach at the leading order. One may simply assume, to estimate theoretical errors, that the uncalculated QCD corrections in $\alpha_s$ are of relative size $\alpha_s(m_V)$ and that the uncalculated corrections in $v$ are of relative size $v^2$.  This leads to ${\cal B}(Z\to J/\Psi J/\Psi) = (1.1\pm 0.4) \times 10^{-10}$ and ${\cal B}(Z\to  \Upsilon(1S) \Upsilon(1S)) = (4.4\pm 0.9) \times 10^{-13}$,  by taking $\alpha_s(m_V)\approx 0.25$ and $v^2 \approx 0.3$  for charmonium, and $\alpha_s(m_V)\approx 0.18$ and $v^2\approx 0.1$ for bottomonium, respectively. These errors are comparable to the ones in eqs. (\ref{rateJtot}) and (\ref{rateUtot}).

A systematical analysis of corrections to our leading-order predictions from higher order $\alpha_s$ and $v$ in the framework of NRQCD will be an interesting theoretical  investigation. Very recently, next-to-leading-order QCD corrections to $Z$-boson decays into double charmonium have been studied and ${\cal B}(Z\to J/\Psi J/\Psi)$ is predicted to be $(1.0\sim 1.3)\times 10^{-10}$ \cite{LFTL22}, which is consistent with our result.

\section{$Z\to V P$}
Similarly, rare decays of $Z\to V P$ can also be analyzed. The leading-order QCD contributions to these processes come from the diagrams like Figure 1, in which one of the $V$'s is replaced by $P$, which have been calculated in Refs. \cite{BR90, LL18}. The corresponding decay amplitude can be written as
\beq\label{amp-VP1}
{\cal M}_1^{VP}=i\frac{512\pi g_v^Q g \alpha_s\psi_V(0)\psi_P(0)}{3\cos\theta_W m_Z^4}\varepsilon^{\alpha\beta\mu\nu}\epsilon^*_\alpha(p)\epsilon^Z_\beta(k)p_\mu q_\nu,
\eeq
where $g_v^Q=T_3^Q-2 e_Q \sin^2\theta_W$ is the vector coupling of $ZQ\bar{Q}$ vertex, and the following projector
\beq\label{projectoretac}
{v(p_{\bar{Q}})} {\bar{u}(p_Q)} \longrightarrow \frac{\psi_{P}(0) I_c}{2\sqrt{3 m_{P}}} i\gamma_5(p\!\!\!/+m_{P})
\eeq
for the pseudoscalar quarkonium $P$ has been used in the derivation. As a good approximation, we set $m_V=m_P$. On the other hand, like the $VV$ final states, as shown in Figure 2, these channels can also happen through $Z\to P\gamma^*$ with $\gamma^*\to V$, which gives
\beq\label{amp-VP2}
{\cal M}_2^{VP}=-i\frac{96\pi g_v^Q g \alpha_{\rm em} e_Q^2\psi_V(0)\psi_P(0)}{\cos\theta_W m_Z^2m_V^2}\varepsilon^{\alpha\beta\mu\nu}\epsilon^*_\alpha(p)\epsilon^Z_\beta(k)p_\mu q_\nu.
\eeq
One can directly find that, now the ratio of two amplitudes ${\cal M}_2^{VP}$ and ${\cal M}_1^{VP}$ is only half of the value for the $Z\to VV$ decays. If we only consider the leading-order QCD contributions, i.e. the amplitude ${\cal M}_1^{VP}$, we have
\beq\label{brJQCD-VP}{\cal B}(Z\to J/\Psi \eta_c) = 9.1\times 10^{-14}\eeq
and
\beq\label{brYQCD_VP}{\cal B}(Z\to  \Upsilon(1S) \eta_b(1S)) = 4.9 \times 10^{-11},\eeq
where we have assumed that $\psi_V(0)=\psi_P(0)$ in the numerical calculation. When both of the amplitudes are included, we obtain
\beq\label{brJ-VP-tot}{\cal B}(Z\to J/\Psi \eta_c) = (1.5\pm 0.4)\times 10^{-11}\eeq
and
\beq\label{brY-VP-tot}{\cal B}(Z\to  \Upsilon(1S) \eta_b(1S)) = (1.9\pm 0.2) \times 10^{-11},\eeq
where the errors are also only due to the uncertainties of $\psi_V(0)$'s.

\section{Summary}

We have presented a theoretical analysis of rare $Z$-boson decays into double heavy quarkonia in the SM. Our study explicitly shows that, besides the leading-order QCD diagrams, another transitions via $Z\to V\gamma^*$ and $Z\to P\gamma^*$, followed by $\gamma^*\to V$, could also bring about significant contributions to these processes. In order to provide up-to-date theoretical predictions for these rare $Z$-boson decay for use in LHC or other future high-precision experimental facilities, we calculate both of them in the present paper. The branching fractions for these decays are predicted, as shown in eqs. (\ref{rateJtot}),(\ref{rateUtot}), (\ref{brJ-VP-tot}), and (\ref{brY-VP-tot}), respectively, which are far below current experimental limits, reported by the CMS Collaboration \cite{CMS22}. In general, it will be challenging to search for such rare processes. However, some interesting room for new physics may be expected in these decays. One could directly utilize some non-standard $ZQ\bar{Q}$ interactions, which, for instance, have been analyzed in Refs. \cite{HV02, GGW16, DSYY22}. The novel couplings might give rise to possible deviations from the SM predictions. Nevertheless, a careful investigation is definitely needed in order to construct some realistic and significant models. This topic is meaningful for the future study.

The enormous events of $Z$ bosons will be produced in the high-luminosity LHC \cite{GKN15, HLLHC19}, or other future experiments such as FCC-ee \cite{FCC-ee} and CEPC \cite{CEPC}, both of which will be planned to run at the $Z$ mass region for a period time. Particularly, at CEPC, running as both a Higgs factory and a $Z$ factory,  a huge number of $Z$ bosons, about ${\cal O} (10^{12})$, would be accumulated.  We look forward to more interesting searches for rare $Z$-boson decays being performed at these machines.

\vspace{0.5cm}
\section*{Acknowledgements}
This work was supported in part by the National Natural Science Foundation of China under Grants No. 11575175 and No. 12047502, and by National Research and development Program of China under Contract No. 2020YFA0406400.


\begin{thebibliography}{40}
\bibitem{CMS19}CMS Collaboration, A.M. Sirunyan et al., Phys. Lett. B {\bf 797}, 134811 (2019), arXiv:1905.10408 [hep-ex].
\bibitem{CMS22}CMS Collaboration, arXiv:2206.03525 [hep-ex].
\bibitem{BR90}L. Bergstr\"om and R.W. Robinett, Phys. Rev. D {\bf 41}, 3513 (1990).
\bibitem{LL18}A.K. Likhoded and A.V. Luchinsky, Mod. Phys. Lett. A {\bf 33}, 1850078 (2018), arXiv:1712.03108 [hep-ph].
\bibitem{colorsinglet}T. Appelquist and H. Politzer, Phys. Rev. Lett. {\bf 34}, 43 (1975); A. De Rujula and S.L. Glashow, Phys. Rev. Lett. {\bf 34}, 46 (1975);
J.H. K\"uhn, J. Kaplan, and E. Safiani, Nucl. Phys. B {\bf 157}, 125 (1979); C. H. Chang, Nucl. Phys. B {\bf 172}, 425 (1980); W.Y. Keung, Phys. Rev. D {\bf 23}, 2072 (1981); E.L. Berger and D. Jones, Phys. Rev. D {\bf 23}, 1521 (1981); L. Clavelli, Phys. Rev. D {\bf 26}, 1610 (1982); L. Clavelli, T. Gajdosik, and I. Perevalova, Phys. Lett. B {\bf 523}, 249 (2001), hep-ph/0110076; L. Clavelli, P. Coulter, and T. Gaidosik, Phys. Lett. B {\bf 526}, 360 (2002), hep-ph/0111250.
\bibitem{barger}V. Barger and R. Phillips, {\it Collider Physics} (updated edition), Westview Press (1996).
\bibitem{Jia2007}G. Hao, C.F. Qiao, P. Sun, and Y. Jia, J. High Energy Phys. {\bf 02} (2007) 057, hep-ph/0612173.
\bibitem{Landau-Yang} L.D. Landau, Dokl. Akad. Nauk SSSR {\bf 60}, 207 (1948); C.N. Yang, Phys. Rev. {\bf 77}, 242 (1950).
\bibitem{BCKLY08}G.T. Bodwin, H.S. Chung, D. Kang, J. Lee, and C. Yu, Phys. Rev. D {\bf 77}, 094017 (2008), arXiv:0710.0994 [hep-ph].
\bibitem{CLY11}H.S. Chung, J. Lee, and C. Yu, Phys. Lett. B {\bf 697}, 48 (2011), arXiv:1011.1554 [hep-ph].
\bibitem{BPSV13}G. Bodwin, F. Petriello, S. Stoynev, and M. Velasco, Phys. Rev. D {\bf 88}, 053003 (2013), arXiv:1306.5770 [hep-ph].
\bibitem{Gao}D.N. Gao, Phys. Lett. B {\bf 737}, 366 (2014), arXiv:1406.7102 [hep-ph].
\bibitem{GG22}D.N. Gao and X. Gong, Phys. Lett. B {\bf 832}, 137243 (2022), arXiv:2203.00514 [hep-ph].
\bibitem{Zpsill}L. Bergstr\"om and R.W. Robinett, Phys. Lett. B {\bf 245}, 249 (1990); S. Fleming, Phys. Rev. D {\bf 48}, R1914 (1993), hep-ph/9304270; S. Fleming, Phys. Rev. D {\bf 50}, 5808 (1994), hep-ph/9403396; CMS Collaboration, A.M. Sirunyan et al., Phys. Rev. Lett. {\bf 121 }, 141801 (2018), arXiv:1806.04213 [hep-ex].
\bibitem{GKPR80}G. Guberina, J.H. K\"uhn, R.D. Peccei, and R. R\"uckl, Nucl. Phys. {\bf B174}, 317 (1980).
\bibitem{BBL95}G.T. Bodwin, E. Braaten, and G.P. Lepage, Phys. Rev. D {\bf 51}, 1125 (1995); Phys. Rev. D {\bf 55}, 5853(E) (1997).
\bibitem{LFTL22}X. Luo, H.B. Fu, H.J. Tian, and C. Li, arXiv:2209.08802 [hep-ph].
\bibitem{GKN15}Y. Grossman, M. K\"onig, and M. Neubert, J. High Energy Phys. {\bf 04} (2015) 101, arXiv:1501.06569 [hep-ph].
\bibitem{HLLHC19}M. Cepeda et al., CERN Yellow Rep. Monogr. {\bf 7}, 221 (2019), arXiv:1902.00134 [hep-ph].
\bibitem{FCC-ee}FCC Collaboration, A. Abada et al., Eur. Phys. J. C {\bf 79}, 474 (2019).
\bibitem{CEPC}CEPC Study Group, J. B. Guimar\~aes da Costa et al., arXiv:1811.10545 [hep-ex].
\bibitem{HV02}X.G. He and G. Valencia, Phys. Rev. D {\bf 66}, 013004 (2002).
\bibitem{GGW16}S. Gori, J. Gu, and L.-T. Wang, J. High Energy Phys.  {\bf 04} (2016) 062, arXiv:1508.07010 [hep-ph].
\bibitem{DSYY22}H. Dong, P. Sun, B. Yan, and C.-P. Yuan, Phys. Lett. B {\bf 829}, 137076 (2022), arXiv:2201.11635 [hep-ph].
\end{thebibliography}
\end{document}